\documentclass{aa}  
\usepackage{tikz}
\usetikzlibrary{shapes.geometric, arrows, positioning}
\tikzstyle{startstop} = [rectangle, rounded corners, minimum width=3cm, minimum height=1cm,text centered, draw=black, fill=blue!25]
\tikzstyle{process} = [rectangle, rounded corners, minimum width=3cm, minimum height=1cm, text centered, draw=black, fill=yellow!30, align=center]
\tikzstyle{decision} = [rectangle, rounded corners, minimum width=3cm, minimum height=1cm, text centered, draw=black, fill=red!30, align=center]
\tikzstyle{arrow} = [thick,->,>=stealth]
\usepackage{graphicx}
\usepackage{physics}
\usepackage{amsmath}
\usepackage{float} 
\usepackage{txfonts}
\usepackage[colorlinks=true,citecolor=blue,urlcolor=blue]{hyperref}

\begin{document}

   \title{Exploring substructures in the Milky Way halo}
   \subtitle{Neural networks applied to {\it Gaia} and APOGEE {\sc dr17}}

   \author{L. Berni\inst{1,} \inst{2},
          L. Spina\inst{1, 3},
          L. Magrini\inst{1},
          D. Massari\inst{4},
          J. Schiappacasse-Ulloa\inst{1}
           \and
          R. E. Giribaldi\inst{1}
          }

   \institute{INAF - Osservatorio Astrofisico di Arcetri, Largo E. Fermi 5, 50125, Firenze, Italy\
         \and
             Dipartimento di Fisica e Astronomia, Università degli Studi di Firenze, Via Sansone 1, 50019, Sesto Fiorentino, Italy\\
            \email{leda.berni@unifi.it}
        \and
             INAF - Padova Observatory, Vicolo dell’Osservatorio 5, 35122 Padova, Italy
        \and
             INAF - Osservatorio di Astrofisica e Scienza dello Spazio di Bologna, Via Gobetti 93/3, I-40129 Bologna, Italy}

  \abstract
   {The identification of stellar structures in the Galactic halo, including stellar streams and merger remnants, often relies on the dynamics of their constituent stars. However, this approach has limitations due to the complex dynamical interactions between these structures and their environment. Perturbations such as tidal forces exerted by the Milky Way, the potential escape of stars, and passages through the Galactic plane can result in the loss of dynamical coherence of stars in these structures. Consequently, relying solely on dynamics may be insufficient for detecting such disrupted or dispersed remnants.}
   {We combine chemistry and dynamics, integrated through a system of neural networks, to develop a clustering method for identifying accreted structures in the Galactic halo.}
   {We developed an integrated approach combining Siamese neural networks (SNNs), graph neural networks (GNNs), autoencoders, and the {\sc Optics} algorithm to create a comprehensive procedure named {\sc Creek}. This method is designed to uncover stellar structures in the Galactic halo. Initially, {\sc Creek} was trained on known globular clusters (GCs) and then applied to the dataset to identify stellar streams.}
   {{\sc Creek} successfully recovered 80\% of the GCs present in the APOGEE dataset, re-identified several known stellar streams, and identified a potential new stream. Additionally, within highly populated stellar structures, {\sc Creek} can identify substructures that exhibit distinct chemical compositions and orbital energies. This approach provides an objective data-driven method for selecting stars associated with streams and stellar structures in general.}
   {}
\titlerunning{Neural networks applied to {\it Gaia} and APOGEE {\sc dr17}}
\authorrunning{Berni et al.}

   \maketitle

\section{Introduction} \label{Section1}

 According to the $\Lambda$ cold dark matter cosmological scenario, galaxies grow their stellar mass, on the one hand, by internal mechanisms such as star formation and, on the other hand, by merger events and accretion of already formed stars in dwarf galaxies or stellar clusters \citep[see e.g.][]{searle1978,bullock2005tracing, cooper2010galactic}. This growth reflects the hierarchical process of structure formation, which also governs the evolution of the Milky Way \citep[MW; see e.g.][]{Rodrigues2017MNRAS.465.1157R, Dai2018ApJ...858...73D, Graaff2024A&A...684A..87D}.
   Stars born from molecular clouds inside the MW form the `in situ' population, while stars accreted from nearby satellite galaxies orbiting the MW create the `ex situ', or accreted population. The presence of different chemical abundance patterns within Galactic halo stars reveals their different origins \citep{Tolstoy2009ARA&A..47..371T, Horta2023, Schiappacasse2025}.
   These dwarf galaxies and stellar clusters are subjected to perturbations such as passages through the Galactic plane or tidal forces caused by the gravitational attraction of the MW. Perturbations can cause a loss of stars from the main structures or even complete disruption of them \citep[see e.g.][]{Onghia2010ApJ...709.1138D, Morinaga2019MNRAS.487.2718M}.
   Stellar streams are one possible outcome of interactions between the MW and accreted globular clusters or dwarf satellite galaxies. In the following, we define a stellar stream as a coherent group of stars, typically originating from the tidal disruption of a satellite galaxy or globular cluster, that exhibits clustering in conserved dynamical quantities (e.g. energy and actions) and shares a chemically homogeneous signature consistent with a common progenitor.
   Nearby streams are usually identified through their kinematic properties \citep[see e.g.][]{belokurov2018co, feuillet2020skymapper, lane2022kinematic}, as the integrals of motion of the stars—such as energy and the vertical component of angular momentum, \( L_z \) (in an axisymmetric system)—are expected to be conserved during the disruption of the progenitor system \citep{helmi2000mapping}, at least for the most recent merger events \citep{chen2024galaxy}.
   With time, the tidally stripped stars are incorporated into the MW halo field population.
   Observations have suggested that a considerable fraction of MW halo stars may originate from disrupted globular clusters, potentially ranging from 10\% to 50\% \citep[see e.g.][]{beasley2020}.
   In addition, debris of streams can mix spatially as time passes, causing one single system to possibly be responsible for multiple streams in a given location of the MW. Along these lines, predictions from numerical simulations suggest that a single accretion event can lead to multiple substructures in phase space \citep[e.g.][]{koppelman2020, Mori2024arXiv240113737M}.
   Some streams have been connected to their progenitors, for example a surviving globular cluster \cite[e.g][]{Jordi2010A&A...522A..71J, Martin2022Natur.601...45M} or dwarf galaxy \citep[see e.g. the stream of Sagittarius, ][]{Belokurov2006ApJ...642L.137B, Gibbons2017MNRAS.464..794G}, but most of them have an unknown origin.
   Large compilations of stellar streams and globular clusters have recently been obtained thanks to the observations collected with the {\it Gaia} satellite \citep[see e.g.][]{Malhan2022ApJ...926..107M, Bonaca2025NewAR.10001713B}.

   However, dynamics alone cannot be used to identify stars that have been completely separated from their globular cluster or galaxy of origin. 
   Chemistry, instead can be considered conserved in these kinds of processes, and it can be used to select stars belonging to a specific structure \citep[see e.g.][]{das2020ages, carrillo2022, lane2023stellar, ceccarelli2024walk}. Indeed, stars that originated from the same star cluster or dwarf galaxy should show, in theory, the same abundances or follow the same abundance pattern, whereas stars originating from different environments should present different abundances. The use of chemistry to re-group stars formed together is called chemical tagging \citep{freeman2002new}.
   The MW stellar halo retains memory of its merger history and contains the remnants of both the first locally formed stars and the accreted ones, thanks to its low stellar density and to the presence of the oldest stellar populations.

   In recent years, the search for stellar clusters and streams has grown substantially, as these structures have become essential tools for probing the formation history of the MW \citep[see e.g.][]{bullock2005tracing, tissera2014stellar, Horta2023}. 
   Indeed, the use of chemical properties to separate stellar populations has traditionally been applied to large structures to distinguish between the high- and low-$\alpha$ sequences both in the disc and in the halo \citep[see e.g.][]{Bensby2003A&A...410..527B, nissen2010, hawkins2015using, Griffith2024arXiv241022121G}.
   Some attempts have also been made to identify dispersed globular clusters and open clusters using only chemical information. For example, \citet{martell2010light} investigated the contribution of high-mass globular clusters to the stellar halo of the Galaxy using CN and CH band strengths as proxies for carbon and nitrogen abundances. This approach was later expanded to include nitrogen and aluminium measurements \citep{martell2016chemical}. Chemical tagging techniques have also been applied to open clusters \citep[e.g.][]{blanco2015testing, Spina2022A&A...668A..16S} and faint dwarf galaxies \citep[e.g.][]{aoki2020chemical}.
   On the other hand, many studies aimed at identifying clusters and streams have relied exclusively on positional data, kinematics, or orbital invariants \citep[see e.g.][]{borsato2020identifying, lovdal2022substructure, dodd2023gaia} and often neglected the chemical properties and ages of stellar populations or considered them only as an a posteriori validation \citep{massari2023cluster, aguado2025cluster, ceccarelli2025cluster}.

   The amount of astrophysical data available today is vastly larger than what astronomers had access to just two decades ago, and it is expected to grow exponentially in the coming years. This rapid increase will be driven by the advent of numerous new surveys and next-generation telescopes \citep[see e.g][]{Jong2019Msngr.175....3D, Babusiaux2019arXiv190404907T, Jin2023MNRAS.tmp..715J, WST2024arXiv240305398M}. Methods that rely only on manual evaluation are no longer applicable to these huge datasets due to both the amount of data to be analysed and the time required.
   As a result, machine learning has emerged as a powerful tool for enhancing data analysis procedures, capitalising on the principle that larger datasets generally improve the performance of machine learning algorithms.
   A few recent works in the literature have demonstrated that supervised and unsupervised machine learning methods are powerful tools for separating structures that have been previously identified by manual analyses \citep[see e.g.][] {lovdal2022substructure, da2023exploring, chen2018chemodynamical, dodd2023gaia}, and a similar use of chemical and kinematic information, though relying solely on a clustering algorithm, was presented by \citet{chen2018chemodynamical}.

   In this work, we present a new approach that combines chemical and dynamical information to identify coherent structures in the MW halo. Our goal is to establish a procedure capable of reconstructing the formation and evolutionary history of the Galaxy by identifying remnants of past accretion events. While applicable to current datasets, this approach is especially designed to leverage the capabilities of upcoming surveys, such as those conducted with the Wide Field Spectroscopic Telescope \citep{WST2024arXiv240305398M}.
   The structure of the paper is as follows: After providing context, in Section~\ref{Section2} we detail the preparation of the observational samples, which are drawn from the APOGEE {\sc dr17} survey \citep[Apache Point Observatory Galactic Evolution Experiment;][]{majewski2017apache}, and the method used to calculate stellar orbits. In Section~\ref{Section3}, we present our algorithm, {\sc Creek}, and this is followed by an analysis of the results in Section~\ref{Section4}, where we discuss the method's strengths and limitations. Finally, in Section~\ref{Section5} we provide our summary and conclusions, outlining the future prospects of our work in the context of upcoming large spectroscopic surveys.

\section{The observational dataset} \label{Section2}

The recent observational advancements are propelling astronomy into a big data era. Large stellar spectroscopic surveys, such as APOGEE \citep[Apache Point Observatory Galactic Evolution Experiment;][]{majewski2017apache}, {\it Gaia}-ESO \citep{gilmore2022gaia, randich2022gaia}, GALAH \citep[GALactic Archaeology with HERMES;][]{desilva2015galah}, LAMOST \citep[Large Sky Area Multi-Object Fiber Spectroscopic Telescope;][]{yan2022overview}, combined with data from the {\it Gaia} missions, 
are providing detailed information about the structure, composition, dynamics, and evolution of the MW.
In this section, we describe the selection criteria to obtain our stellar sample of halo stars, and the methodology used to compute their orbits. This analysis combines spectroscopic data from the APOGEE survey with astrometric measurements from {\em Gaia} {\sc DR3}, allowing for a comprehensive kinematic and chemical characterisation of the stars.

\subsection{The APOGEE survey}

The APOGEE survey was conducted with two 300-fibre cryogenic spectrographs that operate at Apache Point Observatory (APO) in New Mexico, United States, and the 2.5 metre Irénée du Pont Telescope of Las Campanas Observatory (LCO) in Atacama de Chile. APOGEE is unique among large spectroscopic surveys because it is able to acquire medium resolution spectra ($R\simeq 22\,500$) in the near-infrared across all Galactic stellar populations. 
Elemental abundances are derived using the APOGEE Stellar Parameters and Chemical Abundance Pipeline \citep[ASPCAP;][]{perez2016aspcap}. The data release utilised in this work, {\sc dr~17}, contains information for approximately 657,000 targets, providing chemical abundances for up to 20 elements.
In this work, we restrict our sample following the recommendations for quality selection provided by \citet{Horta2023}. Our adopted criteria are signal-to-noise ratio (S/N) > 70, effective temperature (T$_{\rm eff}$) in the range 3500-6500 K, STARFLAG=0 and ASPCAPFLAG< 256\footnote{This selection permits the exclusion of stars with issues associated with stellar parameters and fit processing.}. Our selection closely aligns with \citet{Horta2023}, but we adopted a wider range of values for surface gravity (0<log~g<5) with respect to their selection. 

As a first step, we verified that our selection did not introduce trends or offsets in the abundances as a function of stellar parameters. 
In principle, since clusters are homogeneous in many elements \citep[though not all and not in all elements, as is the case for globular clusters, as described, for instance, in][]{carretta2010properties}, we do not expect to observe trends between [X/Fe] and stellar parameters. Any detected correlations could indicate systematic effects, observational biases, or the presence of additional physical processes influencing the chemical composition of the stars. 
Accordingly, only elements with trends consistent with zero within errors were selected.
In general, we find only negligible correlations between the abundances of elements and either effective temperature or log g with the exception of the member stars in $\Omega$~Cen, which is a peculiar cluster with a complex star formation history \citep[see e.g.][]{carraro2000formation, Tsujimoto2003ApJ...590..803T, Bekki2019ApJ...886..121B}.
Contrary to most abundance ratios, we observe a strong correlation between [C/Fe] and [N/Fe] and log~g. This effect is expected because our sample primarily consists of red giant stars at different evolutionary stages, ranging from the red giant branch (RGB) to the red clump (RC). In the red giant phase, carbon undergoes various mixing episodes, called dredge-ups, which can change its photospheric abundance \citep[see e.g.][]{Lagarde2019A&A...621A..24L}. 

After applying the quality flag to the APOGEE sample, we computed the stellar orbits of our sub-sample.
The orbits were computed assuming the McMillan potential \citep{mcmillan2016mass}, by means of the code {\sc AGAMA} \citep{vasiliev2019agama} using the prescriptions described in \citet{massari2019origin}.
{\sc AGAMA} requires as inputs the right ascension (RA), declination (Dec), distance, radial velocity (with its error), proper motion components in right ascension and declination (with errors) and the correlation coefficient for errors in the two proper motion components. The input data are provided by cross-matching our APOGEE sample with {\it Gaia} {\sc DR3}, where distances are those computed in \citet{bailer2021estimating}. {\sc AGAMA} provides as outputs the apocentre and pericentre of the orbit, the total energy, the angular momentum components, the actions, the circularity and the velocities of stars in the Galactic coordinate system.
The code also creates multiple Monte Carlo realisations of the positions and velocities of each star by sampling from their measured uncertainties to provide uncertainties for the orbital quantities, even though they are not directly used in the analysis. Due to the coordinate system considered to compute orbits in this work, prograde stars are situated in the positive angular momentum side in the Lindblad diagram. Consequently, disc stars are positioned on the right branch (see the right panel of Fig.~\ref{fig:ClusterPanel}).

\subsection{Our sample of halo stars}

\begin{figure*}
\centering
\includegraphics[width=0.9\textwidth]{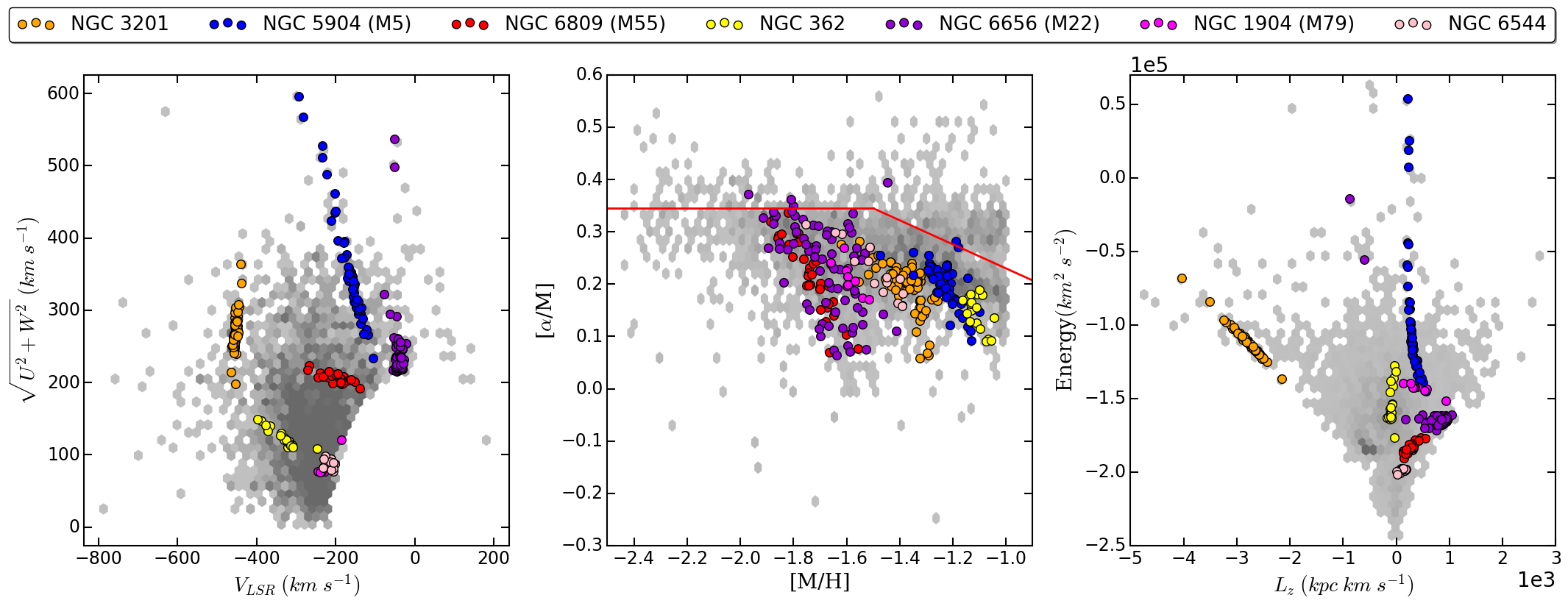}
\caption{\label{fig:ClusterPanel} Left Panel: Toomre diagram for the selected halo stars in APOGEE. Central Panel: [$\alpha$/M] versus [M/H] plane. Right Panel: Lindblad diagram. In all three panels, members of globular clusters from literature (\citet{vasiliev2021gaia}) are colour coded according to the legend, field halo stars are in grey.}
\end{figure*}

Our aim is to select a sub-sample of halo stars, which contains both stars in globular clusters and in the field. We used both chemical and kinematic information to select them. 
Halo stars were selected based on their total velocities $\abs{v_{tot}} > 220$ km~s$^{-1}$ \citep[see e.g.][]{Helmi2017A&A...598A..58H} 
and on their metallicities [Fe/H] < -1 \citep[see e.g.][]{Conroy2019ApJ...887..237C}. 
We also restricted the sample by excluding stars belonging to the Magellanic Clouds with a cut in declination and right ascension. 
Our sample is composed of 3548 stars. The results of our selection are shown in the Toomre diagram, in the left panel of Fig.~\ref{fig:ClusterPanel}.
In this plot, we can recognise the kinematic cut executed to select halo stars. The selection criteria we have chosen are rather tight, because we preferred to avoid the thick disc contamination that would occur, for example, by including total velocities between 150 and 220 km~s$^{-1}$, or by extending metallicity -1$<$[Fe/H]$<$-0.5. In this way, we certainly lose some halo stars, but avoid including a large number of disc stars. 

Of the selected 3548 halo stars, 916 stars belong to 28 different globular clusters, according to the memberships from \citet{vasiliev2021gaia}. Globular clusters with fewer than ten members are not considered in our analysis. In practice, their members are considered as field stars, and we do not attempt to recover them. Instead, globular clusters with ten or more stars are shown in Table \ref{tab:tabClusters}. 
Their locations in the kinematic space are highlighted in the right panel of Fig.~\ref{fig:ClusterPanel}. 
From this plot, we can see that stars in the same cluster generally have coherent velocities and they can be easily identified in the Toomre diagram. We note that clusters appear as elongated structures in these plots due to the use of {\it Gaia} parallaxes—corrected following \citet{bailer2021estimating}—to compute distances for cluster members. It is known that in crowded fields, parallax uncertainties in {\it Gaia} data might lead to trailing features in the Toomre and Lindblad diagrams. This issue could be mitigated by adopting a single distance for all members of a given cluster. However, since we aim to apply the method uniformly to both cluster and field stars, we choose to retain the individual distances derived from {\it Gaia} parallaxes for both populations.
In addition, globular clusters correspond to clumps that exhibit relatively homogeneous metallicity, as explained in further detail in Section \ref{propertiesGC} (see central panel of Fig.~\ref{fig:ClusterPanel}). 
This result highlights that globular clusters are both chemically homogeneous—at least in terms of [Fe/H]—and kinematically coherent. 

\begin{table}[H]
    \centering
    \small
\caption{Globular clusters with more than ten members in our sample, their respective number of selected stars, their mean metallicity, and their origin \citealp[according to][]{ massari2019origin}.}
\begin{tabular}{l l l l}
 \hline
 \hline
Cluster Name & Members & Mean [M/H] & Origin\\ \hline
NGC 5139 $\Omega$Cen &    558 & -1.63& accreted\\ 
NGC 6656 (M22) &     77& -1.68 & in situ \\ 
NGC 3201 &     61 & -1.38 & accreted\\ 
NGC 5904 (M5) &  49 & -1.23 & accreted\\ 
NGC 6809 (M55) &    35 & -1.73 & accreted\\ 
NGC 362 & 20 & -1.11 & accreted\\ 
NGC 6544 & 15& -1.50 & accreted\\ 
NGC 6205 (M13) &   14& -1.50 & accreted\\ 
NGC 1904 (M79) &     10& -1.59&accreted\\ 
NGC 6121 (M4) &    10&-1.04 & accreted\\ \hline 
\end{tabular}
    \label{tab:tabClusters}
\end{table}

\subsubsection{Properties of our sample halo field stars} 

We analyse the chemical properties of our halo sample using their distribution in the [X/Fe] abundance ratios versus [Fe/H] plane.
In some of these planes, in particular those involving $\alpha$ elements, halo stars are separated into two branches. For example, at a given metallicity, two sequences emerge: one corresponding to stars with a higher [$\alpha$/M] abundance (high-$\alpha$ sequence) and another with a lower [$\alpha$/M] abundance (low-$\alpha$ sequence). This separation is particularly evident in the [Mg/Fe] versus [Fe/H] plane, where it has been used by \citet{Hayes2018} to differentiate `in situ' stars from the accreted ones. Stars in their Low-Mg sequence (LMg) generally show halo-like kinematics with little rotation and large velocity dispersion. They are likely accreted stars from other galaxies. On the other hand, stars in their high-Mg sequence (HMg) that show a small velocity dispersion are likely born `in situ', i.e. originated within the MW. There can be some contamination between the two groups, though, as the HMg might contain stars with halo-like orbits. This overlap can be due also to a partial chemical overlap between the two populations.
In the central panel of Fig.~\ref{fig:ClusterPanel}, we plot our sample halo stars in the [$\alpha$/M] versus [M/H] plane\footnote{[M/H] is the total metallicity, and in FGK stars is often approximated with [Fe/H]. In this plot we use [M/H] instead of [Fe/H] because it is directly related to the global [$\alpha$/M].}. The red line marks the separation between the high-$\alpha$ sequence and the low-$\alpha$ sequence. 
The separation between low- and high-$\alpha$ sequences is visible also in the other abundance planes, in particular those involving elements with different production timescales with respect to iron, such as r-process or $\alpha$-elements.

\subsubsection{Properties of our sample globular clusters} \label{propertiesGC}

Most of the globular clusters in our sample are located in the low-$\alpha$ (and correspondingly LMg) region, pointing to their possible origin as accreted objects (see also the classification of \citet{massari2023cluster} reported in Table~\ref{tab:tabClusters}).
In the central panel of Fig.~\ref{fig:ClusterPanel}, we plotted globular clusters on the plane [$\alpha$/M] versus [M/H]. The extragalactic origin of the metal-poor branch clusters is indeed reinforced (or at least not excluded) by their lower [$\alpha$/M] abundances \citep[see. e.g.][]{Recio2018A&A...620A.194R, Belokurov2024MNRAS.528.3198B, massari2019origin}.
As we have already mentioned, $\Omega$ Cen is an anomalous cluster, more similar to a dwarf galaxy than a globular cluster. This also appears in the wide range in metallicity and abundance ratios that its stars cover \citep[see e.g.][]{Johnson2010ApJ...722.1373J, Villanova2014ApJ...791..107V} and we consequently decided not to show it in the plots. 

As shown in the Toomre diagram, clusters can be easily identified in the velocity space: in the left panel of Fig.~\ref{fig:ClusterPanel} we see that they are well grouped in the kinematics space.
We can also question the possibility of distinguishing clusters by using individual abundances instead of metallicity for some representative clusters of our sample. 
The behaviour is not the same for all elements. In fact, we should remember that for some elements, in particular O, Na, Al stars in globular clusters show a large dispersion in abundance. These abundance ratios show anticorrelations, due to the presence of multiple populations in globular clusters \citep{Gratton2012A&ARv..20...50G, Pancino2017A&A...601A.112P}. 
We find that for elements not affected by observational spreads due to the quality of the data and anticorrelations (e.g. Ca, Fe, Ti, Ni), member stars are quite homogeneous and well-separated from the other clusters. By using kinematic and some specific chemical properties, it might be possible, then, to distinguish clusters among them and from field stars.

\subsubsection{Properties of our sample of Galactic halo streams}

Using the members of halo streams selected by \citet{Horta2023} from the APOGEE catalogue, we end up with a list of stream members present in our sample of halo stars. In particular, our sample contains stars belonging to Gaia Enceladus-Sausage (GES), Thamnos, the Helmi Stream, Heracles (Kraken), Arjuna, Wukong, I'itoi, Nyx, Sagittarius, Icarus and three different selections of Sequoia.

\begin{figure*}
    \centering
    \includegraphics[width=0.9\linewidth]{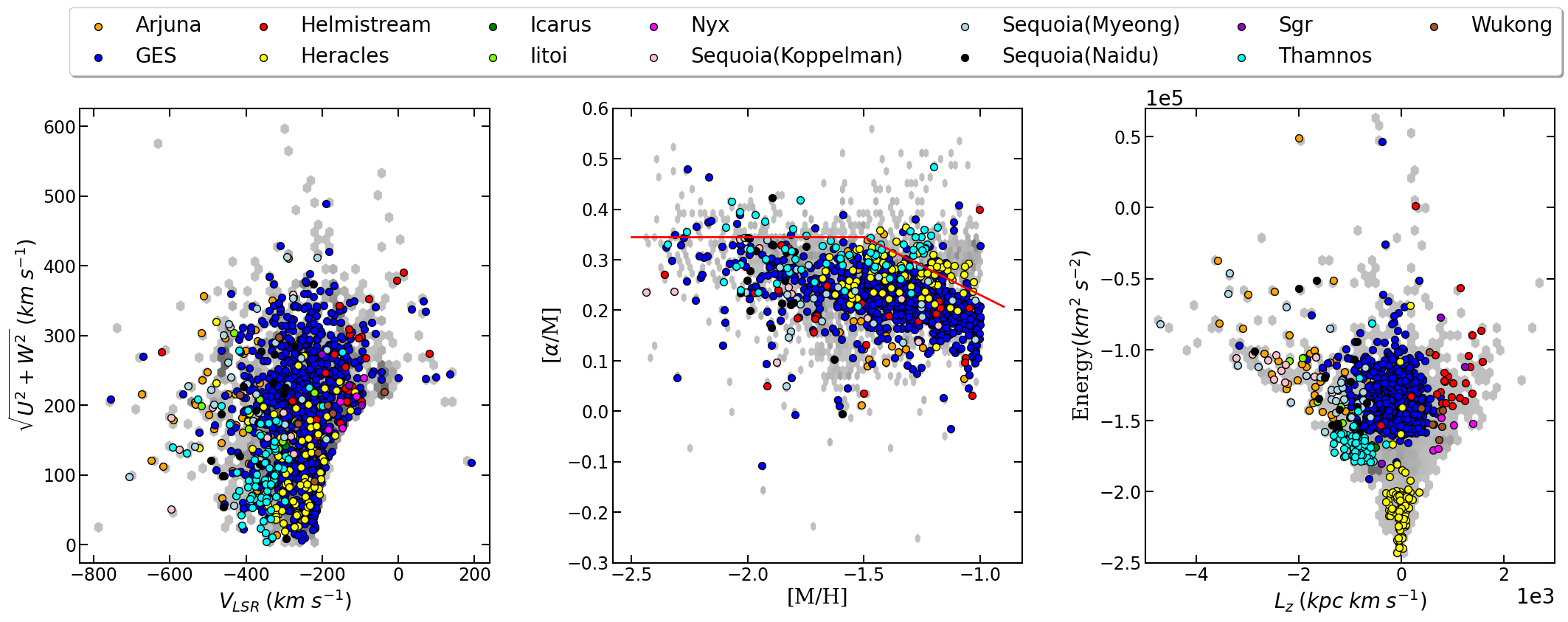}
    \caption{Left Panel: Toomre diagram. Central Panel: [$\alpha$/M] versus [M/H] plane. Right Panel: Lindblad diagram. Prograde streams are positioned on positive values of $L_Z$ and retrograde streams are on negative values. In all panels, stars in gray represent field halo stars, colour coded according to their density, stars belonging to streams from \citet{Horta2023} are highlighted according to the legend.}
    \label{fig:StreamsPanel}
\end{figure*}

In the left panel of Fig.~\ref{fig:StreamsPanel}, we plot the velocities of the members of the streams in the Toomre diagram. The result is quite different with respect to that of globular clusters. 
Globular clusters are clearly identifiable and grouped. Streams, instead, are sparse and spread over the plane, even when excluding stars from the largest merger, GES, one of the most important known streams in the solar neighbourhood \citep[see e.g.][]{helmi2018merger}. It represents the dynamical record in the velocity space of a major collision that the MW experienced more than ten Gyr ago with a quite massive dwarf galaxy \citep{belokurov2018co}. Its high spread in velocities, but also -- as we show later -- in chemistry is due to its past as a galaxy, with an extended star formation history and with an elevated star formation rate
\citep{Vincenzo2019}. 
We therefore expect GES to be more diffuse and spatially extended compared to other stellar streams. However, even the other streams identified in APOGEE are challenging to distinguish in the Toomre diagram.
This difficulty arises because, over time, stars originating from past merger events have gradually lost their kinematic coherence due to phase mixing and dynamical interactions within the MW. Nonetheless, they may still preserve certain orbit-invariant quantities, such as the total energy, the vertical component of angular momentum ($L_Z$), or their chemical signatures \citep[e.g.][]{Bonaca2021ApJ...909L..26B}.
  
In the central panel in Fig.~\ref{fig:StreamsPanel}, we show the abundance ratio [$\alpha$/M] versus metallicity [M/H].
Stream stars generally show a low [$\alpha$/M] abundance since they are usually an accreted population. For instance, the Sagittarius stream, the Helmi stream, Sequoia and GES are located in the low sequence of [$\alpha$/M], but we find a few notable exceptions in Thamnos and some among Heracles stars that seem to be $\alpha$-enhanced or located at the limit between the two sequences. This might be explained as them being part of the category of stellar substructures created by past interactions in the disc or secular processes. Thamnos was renamed as `Erebus' by \citet{giribaldi2023chronology}, who find that its stellar ages ($\sim$12 Gy) and its [Fe/H] spread to values as high as -0.5 dex indicate its origin is most likely `in situ', despite the fact that the orbits of their stars are retrograde.
In the right panel of Fig.~\ref{fig:StreamsPanel}, we plot the members of the streams in the energy versus $L_Z$ plane.
Most of the streams, such as Sequoia, Thamnos and Arjuna are located in the retrograde zone (to the left) and around $L_Z = 0$, where the broad Gaia-Enceladus remnant is particularly noticeable in the central area of the diagram. Still, there are some exceptions: Nyx and the Helmi stream are on the prograde side of the plot.

\subsection{Selecting the dataset}

In the previous sections, we have seen that dynamics alone is insufficient to fully recover structures, as ancient mergers undergo phase mixing, leading to the loss of dynamical coherence \citep{Mori2024arXiv240113737M}. The chemical composition of stellar photospheres, however, remains preserved over time. Therefore, our search for disrupted structures relies on the most effective chemical tracers, complemented by orbital properties.
For this reason, to identify past mergers and disrupted clusters through their chemical signatures, we must carefully select the most informative elements. Additionally, to assess the effectiveness of our method, we need to test it on the most chemically compact clusters, which provide the clearest benchmarks.

The selection of the most effective elements relies on two key factors: their ability to distinguish populations with different star formation histories and the precision with which their abundances can be measured.
Indeed, stars will be more clustered in some abundances compared to other ones, as different elements come from different nucleosynthesis processes. This is a crucial point, as depending on the observed clusters, we can find that a different element separates them best.
While some elements, such as neutron-capture elements such as Eu \citep[see e.g][]{monty2024ratio}, are theoretically excellent tracers of stellar populations with different origin, their reliable measurement in medium-resolution spectra, such as those from APOGEE, is extremely challenging. Additionally, determining abundances in low-metallicity stars remains particularly difficult, further complicating their use as chemical tracers.

In order to choose the best elements, we based our selection on the abundances of globular clusters, selecting the best elements to identify globular clusters from field halo stars.
This approach follows a metric inspired by \citet{Mitschang2013} to identify the most effective chemical abundances for distinguishing stellar populations. Specifically, we aim to determine the abundances that maximise the separation in chemical space between stars within a given cluster compared to those from different clusters.

To achieve this, we defined the following metric:
\begin{equation} \label{Metric}
    \delta_c = \sum_{i,j}^{i\neq j}{\frac{\abs{A_C^i-A_C^j}}{N_{stars}}},
\end{equation}
where $i$ and $j$ are indices of the stars, $A_C$ are the abundances of the element and $N_{stars}$ is the number of stars considered. We compute $\delta_C$ for stars within the same cluster, yielding an intra-cluster metric, and for stars in different clusters, yielding an inter-cluster metric.
The ratio between intra-cluster and inter-cluster metric gives us the key parameter for assessing the suitability of a given element for clustering analysis. A lower value of this ratio indicates that the element is effective at distinguishing between different clusters while maintaining a compact distribution within individual clusters. Ideally, the best elements for clustering applications should have the following parameter close to zero: 
\[
p = \frac{\delta_C^{intra-cluster}}{\delta_C^{inter-cluster}}.
\]
This metric is extremely sensitive to the number of clusters and to the number of members within the clusters used to compute it. For instance, we get different values if we only consider the most massive clusters or if we include all of them.
After a thorough analysis of the elements available in the APOGEE database, and considering a balance between the value of the metric for each element, the number of stars with a measure of that abundance, and the typical error, we opted to use metallicity and [$\alpha$/M] as tracers for the clusters. This decision was also supported by the fact that Mg, Si, Ca, Ti, Zn and Eu, which are $\alpha$ elements or produced with short time scales, are the most sensitive elements to the origin of globular clusters \citep{ceccarelli2024comparative}.
Metallicity [M/H] and iron abundance [Fe/H] are definitely the parameters with the best metrics approaching 0.2. Among the other elements, [$\alpha$/M] is available in all the stars in the sample and has a parameter value of about 0.68. Other abundance ratios that have metric values similar to [$\alpha$/M] are [Mg/Fe] and [Si/Fe], as might be expected since they are $\alpha$ elements, but they essentially give us the same information and there is no added value to include them.

\section {{\underline C}lustering with g{\underline R}aph n{\underline E}ural n{\underline E}twor{\underline K}: {\sc Creek}}
\label{Section3}

The standard analysis for a chemical clustering consists in applying a clustering algorithm directly to the abundances of the samples of stars. However, as we have seen in Fig. \ref{fig:ClusterPanel} and \ref{fig:StreamsPanel}, both globular clusters and stellar streams can be superimposed one on each other in the chemical space. This is preventing us from retrieving these stellar populations purely from their chemical make up. In fact, recent studies aiming at finding stellar populations in the Galactic halo prefer to identify coherent structures and clumps in the integral of motion and the considering the chemical properties only for a posteriori validation \citep[see e.g.][]{lovdal2022substructure, Ruiz2022A&A...665A..58R}. However, this strategy brings the risk of including many contaminants and losing several true members of these populations. In fact, stellar dynamics could have been modified by encounters with the gravitational potential of the Galaxy, while their chemical pattern is the only invariant in time. 

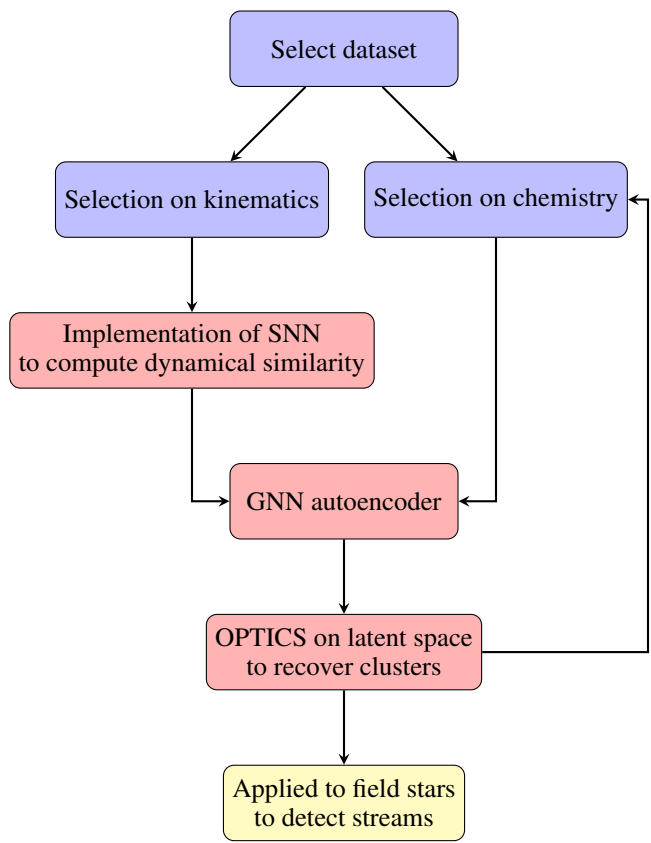
\begin{figure}
    \centering
    \begin{tikzpicture}[node distance=2cm]

        \node (start) [startstop] {Select dataset};
        \node (vel) [startstop, below of=start, xshift=-2cm] {Selection on kinematics};
        \node (chem) [startstop, below of=start, xshift=2cm] {Selection on chemistry};
        \node (snn) [decision, below of=vel] {Implementation of SNN\\ to compute dynamical similarity};
        \node (gnn) [decision, below of=start, yshift=-4cm] {GNN autoencoder};
        \node (optics) [decision, below of=gnn] {OPTICS on latent space\\ to recover clusters};
        \node (streams) [process, below of=optics] {Applied to field stars\\ to detect streams};

        \draw [arrow] (start) -- (vel);
        \draw [arrow] (start) -- (chem);
        \draw [arrow] (vel) -- (snn);
        \draw [arrow] (snn) |- (gnn);
        \draw [arrow] (chem) |- (gnn);
        \draw [arrow] (gnn) -- (optics);
        \draw [arrow] (optics) -- (streams);
        \draw [arrow] (optics) -- ++(4cm,0) |- (chem);

    \end{tikzpicture}
    \caption{Flowchart of our algorithm used for deep clustering.}
    \label{fig:chart}
\end{figure}

Instead, we have developed an algorithm that searches for stellar populations in the halo by making a synergic use of both chemistry and dynamics. However, since chemistry is invariant and dynamics is not, these two pieces of information cannot be considered on the same level. Thus, our strategy comprises identifying these groups mainly based on the chemical space, but it has also been adapted to consider the dynamical similarity between stars. A simple flowchart that represents the final structure of the algorithm used for deep clustering, named {\sc Creek} ( {\underline C}lustering with g{\underline R}aph n{\underline E}ural n{\underline E}twor{\underline K}), is depicted in Fig.~\ref{fig:chart}. The name `{\sc Creek}' also recalls the structure of the streams, which appear as rivers or creeks in the halo of the MW. 

\begin{figure}
    \centering
    \includegraphics[width=0.9\linewidth]{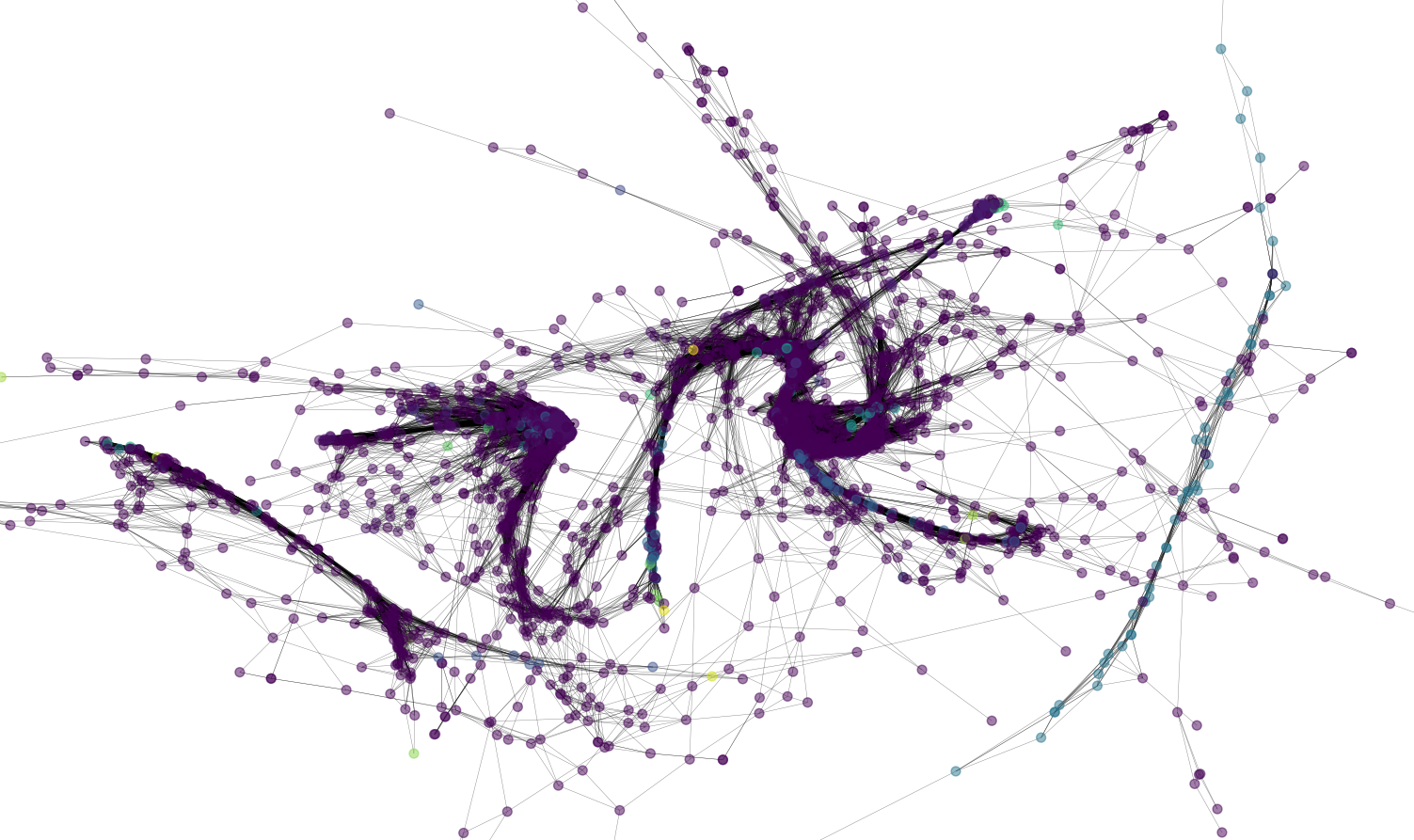}
    \caption{Example of graph obtained in this work. Nodes are represented by dots and coloured according to their cluster from literature. Edges are shown as lines.}
    \label{fig:Graph1}
\end{figure}

First we use the integrals of motion to identify pair of stars that have similar dynamics. Once these stars are identified, they are linked together. The ensemble of these links and the chemical space form a graph. Graphs are composed of nodes and edges, where nodes typically represent entities or objects, and edges capture the relationships or interactions between them. An example of graph is represented in Fig.\ref{fig:Graph1}. The graph is then given as an input to a Graph Neural Network, which aggregates the chemical information of all the stars linked together and creates a latent space on which we perform the clustering analysis using {\sc Optics}. Below we review these three steps of the analysis giving more details on how the dynamical and chemical information is used in order to search for chemo-kinematic structures in the halo of the Galaxy.

\subsection{Establishing the dynamical similarity}
The main goal of this first step is establishing the dynamical similarity between stars. Dynamically similar stars are those linked by edges in the graph. For this task we make use of a Siamese Neural Network \citep[SNN; see e.g.][]{Chicco2021}.

A simple neural network (NN) consists of layers of perceptrons, where each layer is typically connected to the previous and next ones. It takes input features from a dataset and optimises parameters called weights and biases during training to perform a given task. The NN includes an input layer (matching the number of input features), an output layer (matching the predicted values), and hidden layers in between. Each hidden layer applies a transformation using a weight matrix $W$, a bias vector $B$, and a usually non linear activation function $\Phi$, following the equation $Z = \Phi(XW^T+B)$. This process continues through all layers until the final predictions are generated. 

SNNs consist of a neural network architecture with two or more identical sub-networks that share the same structure, parameters, and weights. They process pairs of input instances to determine their similarity or dissimilarity. Each input in a pair is fed through the same network (the 'Siamese twins'), and their outputs are compared using a metric such as Euclidean distance. Their result is then processed by a neuron with a Sigmoid activation function, 
\begin{equation} \label{eq:sigmoid}
    \sigma(x) = \frac{1}{1+e^{-x}},
\end{equation}
producing an output between 0 and 1 that represents the probability that the inputs are identical. During training, this probability is compared to the true labels for parameter optimisation, similar to standard artificial neural networks. Once trained, the SNN can evaluate new input pairs, returning a probability of their similarity.

We used the orbital properties of our stellar sample to implement a SNN, leveraging dynamical information—specifically the standardised actions $J_R$, $J_{\phi}$, and $J_Z$. Here, standardisation refers to the process of subtracting the mean and scaling by inverse standard deviation for each input feature. These quantities are particularly effective for recovering structures such as stellar streams or merger remnants that have lost coherence in position and velocity space, as they serve as conserved integrals of motion, at least to first approximation.

\begin{figure}
    \centering
    \includegraphics[width=0.9\linewidth]{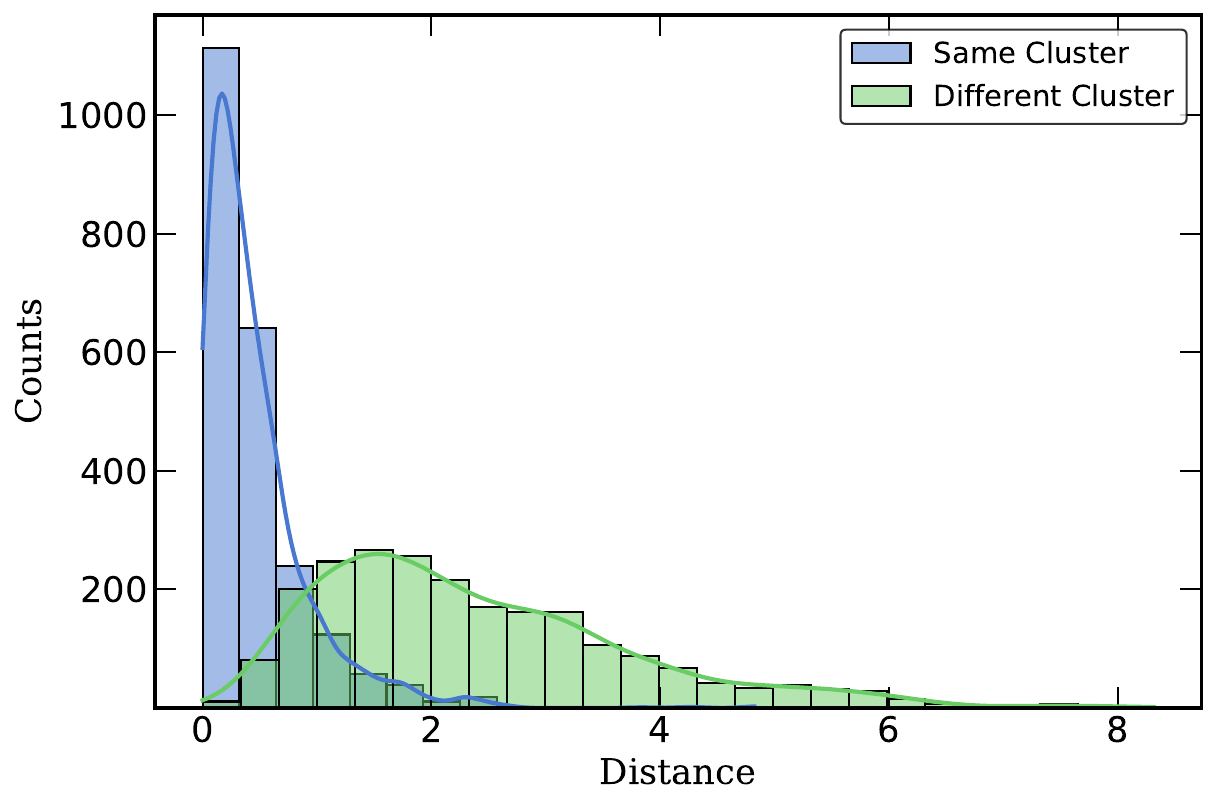}
    \caption{Histogram of the Euclidean distances in the space of standardised orbital actions for the pairs of stars used to train the SNN. Stars belonging to the same cluster are shown in blue, while and stars belonging to different clusters are green.}
    \label{fig:histcluster}
\end{figure}

We implemented our SNN by first constructing a training set using a random selection of stars from globular clusters. We selected 2250 pairs of stars from the same cluster and an equal number of pairs from different clusters. Maintaining this balance ensures a well-structured and unbiased training sample, improving the network's ability to distinguish between similar and dissimilar pairs.
Our SNN took as input the orbital parameters of two stars and was trained to predict whether they belong to the same association. The network consisted of two identical branches, each with three hidden layers containing 32, 64, and 128 neurons, all using the rectified linear unit (ReLU) activation function \citep[see e.g.][]{Arora2016arXiv161101491A}. The dataset was split into 80\% entries for training and 20\% for validation. The two branches shared the same weights and biases, ensuring consistent feature extraction. Their outputs were used to compute a distance metric, which was then passed through a Sigmoid activation function, as shown in eq.\ref{eq:sigmoid}. The SNN uses a binary crossentropy loss and achieves an accuracy on the validation set of 0.99. The final output represented the probability that the two stars belong to the same cluster, approaching 1 for stars from the same cluster and 0 otherwise.
In Fig.~\ref{fig:histcluster} we plotted the two distributions of the kinematic distances in the training set of the SNN. The distribution of intra-cluster stars (light blue) appears compact, indicating greater coherence in their orbital actions. In contrast, inter-cluster stars (green) tend to have larger distances, reflecting their separation across different clusters.
The Sigmoid activation function is particularly useful because it allows for flexibility in setting the threshold for classifying stars as belonging to the same cluster. This adaptability enables different observational goals: lower probability thresholds can be used to identify streams or stars from disrupted clusters, while higher thresholds help isolate the dense cores of clusters.
In this work, we use a threshold value of 0.8 to determine whether stars are linked. Stars with a probability of at least 0.8 of belonging to the same cluster are considered connected and are used to construct the graph for the next neural network (see Section~\ref{sec_GNN}).
Although the SNN is trained only on members of globular clusters, we find that it is able to identify pairs of stars on similar orbits, whether they belong to streams, globular clusters or the field.

\subsection{Creating the latent space}
\label{sec_GNN}

The stellar abundances and the links established through the SNN served as the inputs for the graph neural network (GNN) autoencoder \citep[see. e.g.][]{Gori2005}. 
GNNs operate on graph structures of data, where data are called nodes and links are called edges.
At each layer of a GNN, information from each node is aggregated and combined with that of all nodes connected to it, including itself if it has a self-loop. The process is iterated over multiple layers as in any neural network, allowing nodes to gather and incorporate information from their connections as well as from the broader graph structure. In the present work, aggregation was realised through a sum weighted on the number of edges connected to the node. This was done to balance nodes with varying connectivity, ensuring that those with many edges did not disproportionately influence the neuron's output compared to nodes with few or no edges.

The GNN is also an autoencoder. This type of NN is formed by two parts, one with a progressively decreasing number of neurons per layer, called encoder, and one with a progressively increasing number of neurons per layer, called decoder. The bottleneck of this structure is called latent space. The aim of the autoencoder is to make the output as similar as possible to the input, so that the latent space provides a more efficient representation of the input values, while preserving most of the information.
In this work, the autoencoder mapped the chemical space into a latent space that also incorporated the dynamical similarity between stars in our dataset. Rather than processing simple chemical abundances, the graph autoencoder propagates the averaged information abundances across the entire network. As a result, stars with similar orbital parameters are expected to be located close together in the latent space, making stellar associations denser and easier to identify by clustering algorithms compared to the original chemical space.
To ensure efficient training, the input data was standardised. Without standardisation, uneven gradients in the loss function—which quantifies the discrepancy between the model's predictions and the true values—can slow down the learning process, as the optimiser requires more steps to adjust some dimensions than others. Therefore, we standardised each abundance before passing the data to the autoencoder.
The autoencoder architecture consisted of five hidden layers, with neuron counts varying across different layers. The configurations tested included 500 and 300 neurons for the outer layers, 100, 200, or 300 for the inner layers, and either three or five neurons for the latent space. The final configuration used layers with 300, 100, and 5 neurons and a  mean squared error (MSE) loss. The model was trained for 800 epochs, with 500 and 1000 epochs also tested but deemed less effective due to longer compilation times and poorer results. The autoencoder was equipped with a learning rate scheduler with a `patience' -- the number of epochs to wait before early stop if no progress on the validation set -- of 50 and a reduction factor of 0.5, as well as an early stopping callback with a `patience' of ten. These hyperparameters were chosen based on extensive exploratory tests.
After setting the hyperparameters, we repeated the training multiple times and retained the model with the lowest loss.
Interestingly, even when only two abundances were considered, an autoencoder with a five-dimensional latent space was more effective at recovering clusters than those with smaller latent spaces.

\subsection{Recovering chemo-dynamical structures}
{\sc Optics} was then applied to the latent space of the autoencoder, where the information was simplified. Prior to clustering, the latent space was standardised to ensure that each dimension contributed equally. This step was essential to prevent any single dimension from dominating the clustering process.
{\sc Optics} generated a reachability plot, which visualises the clustering structure by displaying the reachability distance of each data point in the order they were processed\footnote{Reachability distance is the distance between a data point and its nearest core point (a point with sufficient neighbours within a given radius)}. In that plot, valleys or dips indicate dense clusters and peaks represent noise or cluster boundaries.
By comparing dips with the actual clusters in the dataset, we can assess the effectiveness of the clustering.
In the next section, we describe the application of the algorithm to the Galactic halo sample, present the recovery factor for the globular cluster sample, and discuss the method's limitations in recovering known streams from the literature and possibly discovering additional ones.

\section{Results of the analysis: Application of {\sc Creek} to the APOGEE dataset}
\label{Section4}

\subsection{APOGEE}

We applied {\sc Creek} to the APOGEE dataset, first to recover known globular clusters and then to identify mergers and streams while also aiming to possibly find new ones.
{\sc Creek}, with {\sc Optics}, identifies high-density groups of stars within the latent space. Therefore, we can evaluate the performance of the algorithm by comparing these groups with the real stellar associations. More specifically, we defined the metrics `homogeneity' $H$ and `completeness' $C$, which we computed as follows:
\begin{equation}
    H = \frac{Number\;of\;cluster\;stars\;included\;in\;group}{Number\;of\;stars\;in\;group}.
\end{equation}
\begin{equation}
    C = \frac{Number\;of\;cluster\;stars\;included\;in\;group}{Number\;of\;stars\;in\;cluster}
.\end{equation}
Here, `cluster' is the astrophysical entity defined using a membership analysis based on independent quantities, such as proper motions and radial velocities, while `group' is the set of stars with similar properties assembled by {\sc Creek}.
We consider a cluster as recovered if an {\sc Optics} group has both homogeneity and completeness over 30\%: $H\geq 0.3$ and $C\geq0.3$.

\begin{figure*}
    \centering
    \includegraphics[width=0.9\linewidth]{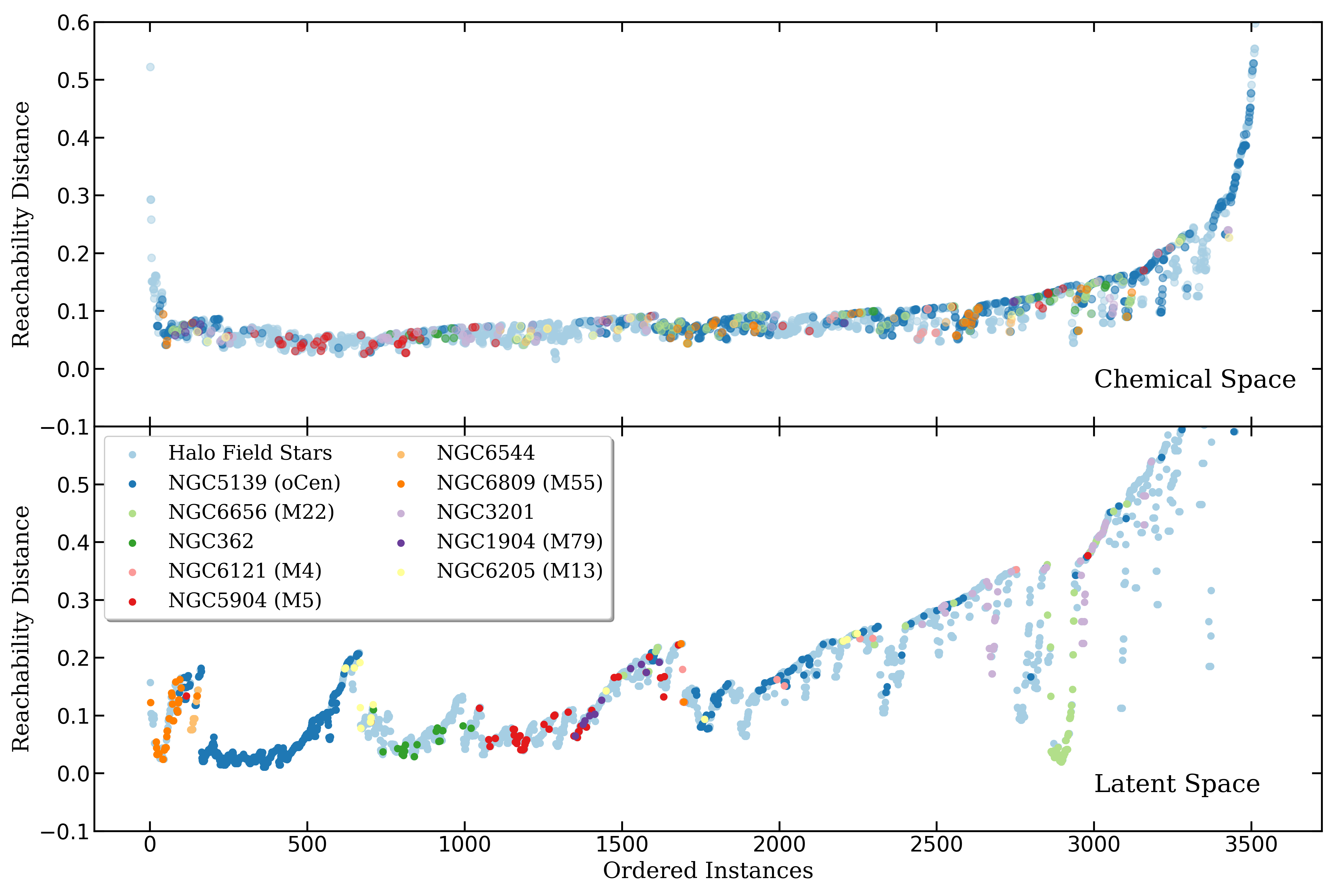}
    \caption{Reachability plots (reachability distances versus ordered instances). In the upper panel, {\sc Optics} was applied directly to the chemical space, and in the lower panel it was applied through {\sc Creek} to the latent space. Light blue dots are field stars, and coloured dots are stars belonging to known clusters.}
    \label{fig:ExReachPlot}
\end{figure*}

In Fig.~\ref{fig:ExReachPlot}, we present an example of a reachability plot. Each dip in the plot represents a potential group, and the colours correspond to clusters as defined in the literature. The upper panel shows the application of {\sc Optics} directly to the dataset containing only chemical abundances, while the lower panel demonstrates the application of {\sc Optics} in the latent space defined by our SNN and GNN.

Even before calculating the number of recovered clusters and evaluating their homogeneity and completeness, we observe that the clusters are better separated and more easily identifiable in the lower panel. The stars that belong to the same cluster tend to be grouped together within the dips of the reachability plot, indicating improved clustering performance when using the latent space representation.

\subsubsection{Globular clusters in APOGEE}

We applied {\sc Creek} algorithm to the APOGEE dataset to test its ability to recover globular clusters. 
Depending on the parameters of {\sc Optics}, the hyperparameters of the autoencoder, and the model considered the number of clusters recovered can vary. In this case, we consider the configuration with {\sc Optics} parameters $\xi = 0.001$ and $\text{min\_sample} = 7$.
A value of $\xi = 0.001$ indicates that the steepness of a dip required to be considered a cluster is relatively small. This means that even slight changes in reachability distance between neighbouring points can be detected as potential clusters. Consequently, the reachability plot shows a more gradual change in the distances, which may lead to identifying smaller or more loosely defined clusters. The $\text{min\_sample}$ parameter sets the minimum number of points required to form a dense region, influencing the sensitivity to smaller or sparser clusters.
The reachability plot obtained for this configuration is shown in Fig.~\ref{fig:ExReachPlot}.
{\sc Optics} produces a long hierarchy of groups and sub-groups, eight of which correspond within the $H$ and $C$ criteria established above to real clusters. The recovered clusters are in Table \ref{tab:Clrec05}.

\begin{table}
\caption{Recovered globular clusters.}
    \centering
    \tiny
    \begin{tabular}{lll} \hline \hline
         Cluster & Homogeneity & Completeness \\ \hline 
         NGC 5139 ($\Omega$Cen) & 0.94 & 0.65 \\  
         NGC 6656 (M22) & 0.67 & 0.77 \\ 
         NGC 6121 (M4) & 0.50 & 0.50 \\ 
         NGC 6544 & 0.54 & 0.93 \\ 
         NGC 362 & 0.32 & 0.30 \\ 
         NGC 6809 (M55) & 0.31 & 0.86 \\ 
         NGC 5904 (M5) & 0.40 & 0.43 \\ 
         NGC 3201 & 0.56 & 0.33 \\ \hline
    \end{tabular}
    \label{tab:Clrec05}
\end{table}

\begin{figure*}
    \centering
    \includegraphics[width=0.9\linewidth]{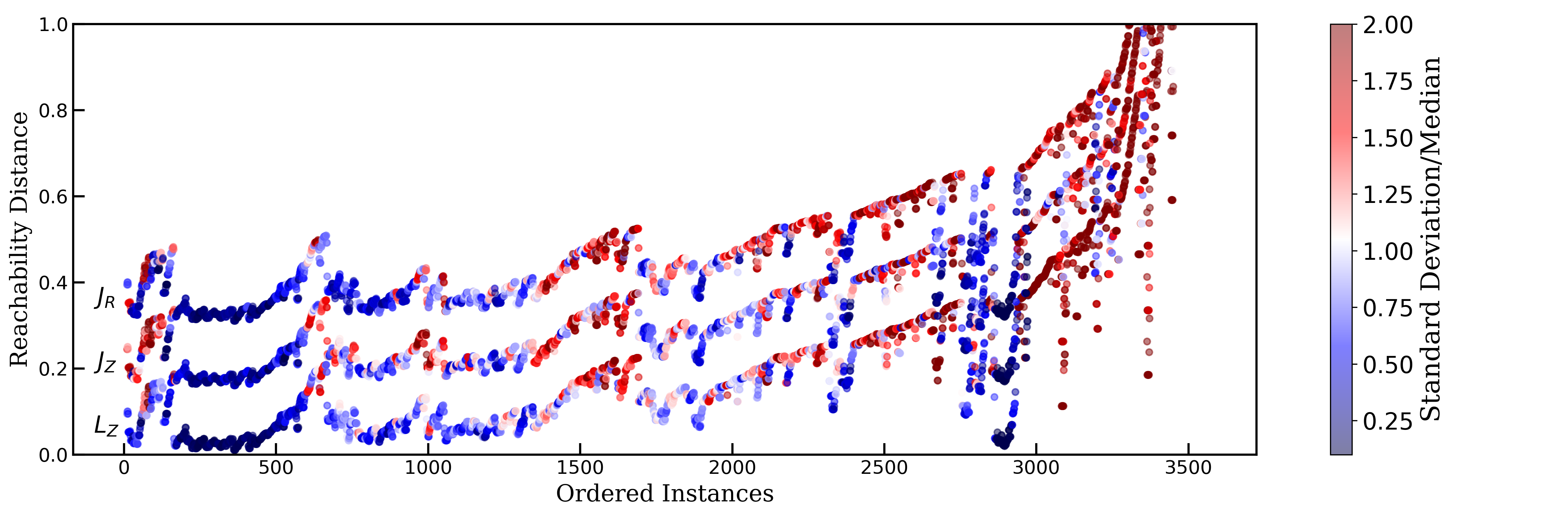}
    \caption{Reachability plot of the APOGEE dataset. The points are 
    colour-coded according to the ratio between the standard deviation and the median over segments of length ten in $L_Z$, $J_R$ and $J_Z$. Blue (red) stars have a lower (higher) standard deviation. The three variables are plotted with a vertical offset to avoid overlapping.}
    \label{fig:std}
\end{figure*}

Another important insight can be gained from Fig.~\ref{fig:std}. In this figure, colours are assigned to instances based on the ratio between the standard deviation and the median over segments of length ten in three different parameters: $L_Z$, $J_R$, and $J_Z$. This visualisation highlights that most dips in the reachability plot correspond to regions with a lower standard deviation in these three variables compared to stars located outside the dips. This suggests that the groups identified by {\sc Creek} correspond to stars with a smaller spread in the action space, indicating they are well clustered in that space, thus confirming the effective retrieval of clusters. However, not all dips exhibit a lower standard deviation, which is expected, as some dips may result from chemical resemblances rather than coherence in orbital actions.

\subsubsection{Stellar streams and merger remnants in APOGEE}

\begin{figure*}
    \centering
    \includegraphics[width=0.9\linewidth]{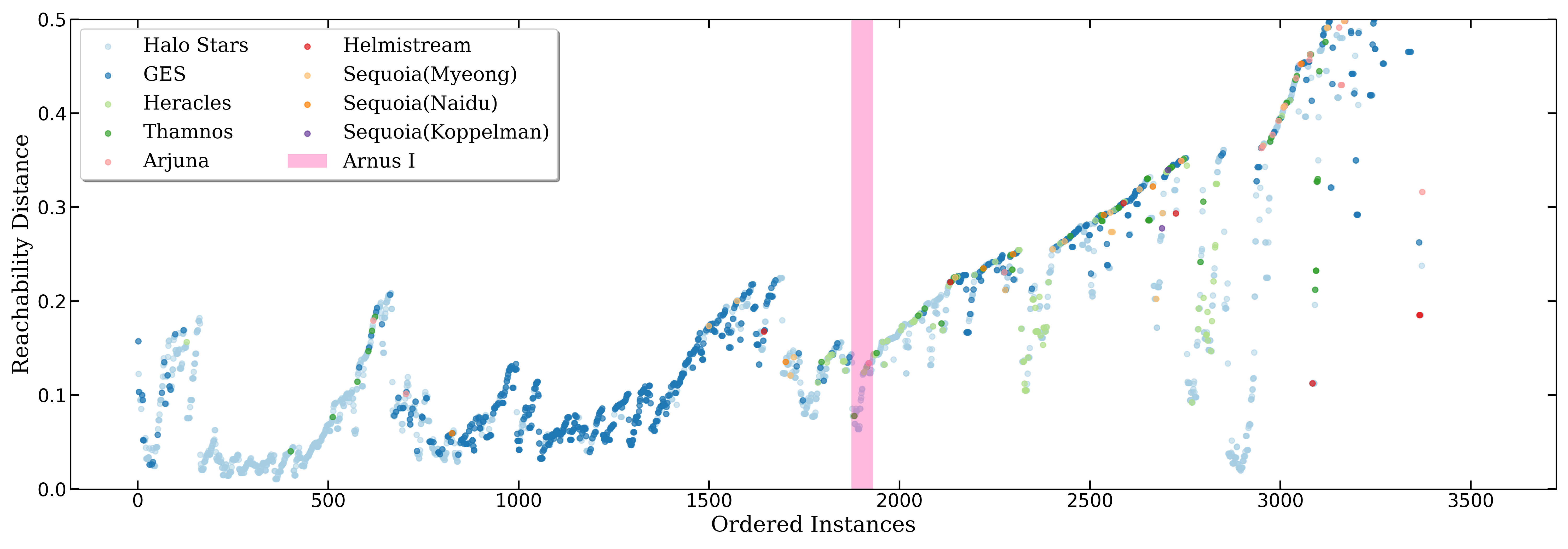}
    \caption{Reachability plot for the stellar streams in the APOGEE dataset. Stream stars are colour-coded as in the legend.}
    \label{fig:stramreach}
\end{figure*}

Since we have shown that our algorithm is able to recover 80\% of clusters to at least 30\% in completeness and homogeneity, we can now focus on identifying non-spatially coherent structures, such as stellar streams and merger remnants.
We used the list of stream members defined according to the selection criteria of \citet{Horta2023}.
We aim to investigate the placement of well-studied streams in the reachability plot and determine whether they coincide with dips.
Since streams may not be entirely chemically homogeneous, having likely formed from galaxies or clusters with an extended star formation history, we avoided computing the effective recovery factors (as we did for globular clusters). We coloured their locations in the reachability plot as shown in Fig.~\ref{fig:stramreach}. Although sparser than clusters, stream stars are still positioned within the dips of the reachability plot and can be distinguished from globular clusters. In fact, they occupy distinct dips compared to globular clusters in the plot.
In what follows, we describe some of the most interesting streams (re)-identified with {\sc Creek}.

\paragraph{Gaia-Enceladus}

\begin{figure*}
   \centering
   \includegraphics[width=0.9\linewidth]{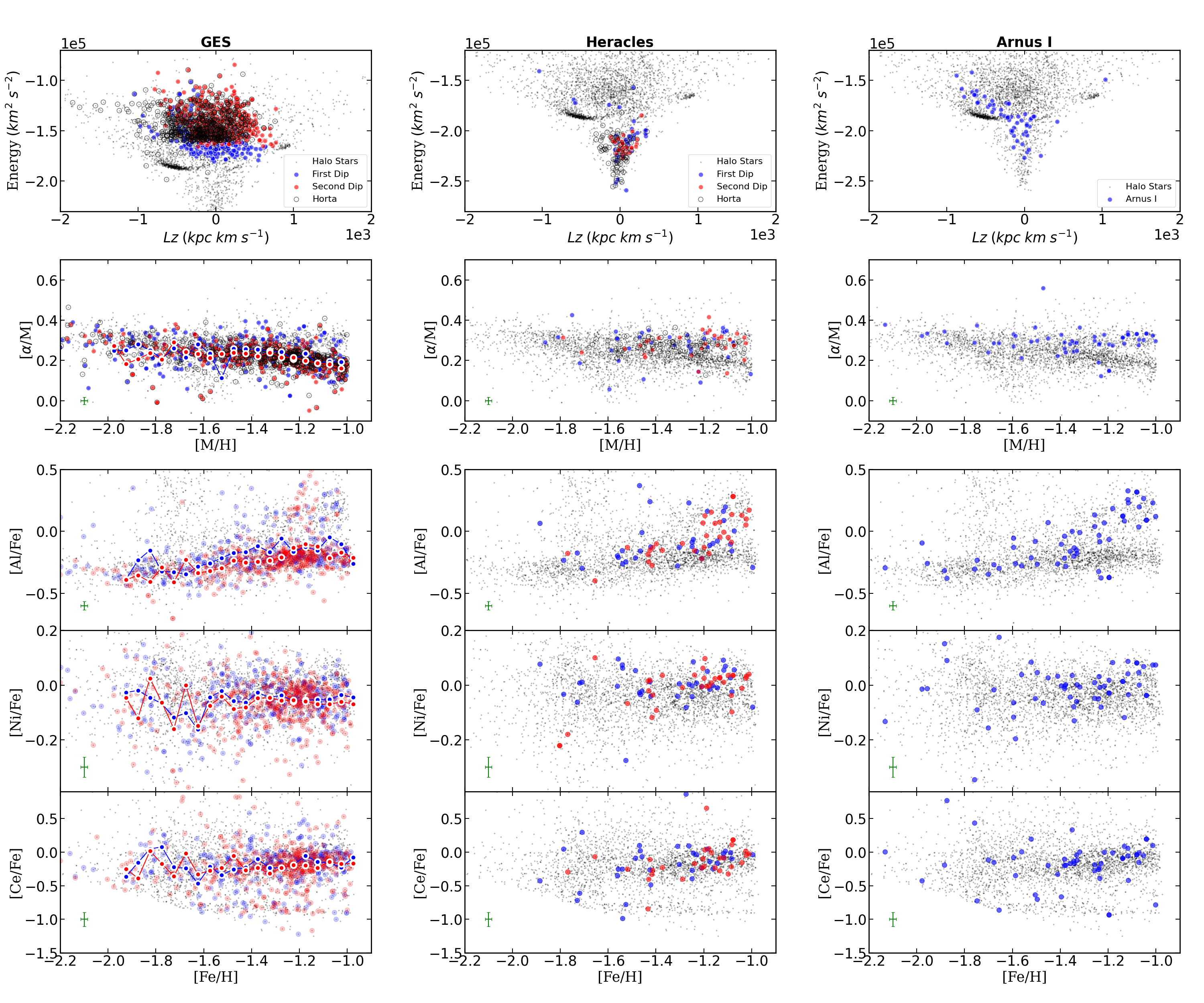}
   \caption{First column: Lindblad diagram,[$\alpha$/M] versus [M/H], and abundance ratios of Al, Ni, and Ce over Fe as a function of [Fe/H] for stars in GES. The halo sample is represented with grey dots, the GES first dip population is shown with semi-transparent blue circles, and the second dip is in red. Mean abundance of the $\alpha$-elements, [Al/Fe], [Ni/Fe], and [Ce/Fe] of the first and the second dip of GES are shown with blue and red circles, connected by continuous lines. The green crosses represent the mean error on the element. Second Column: the same plots are repeated for the Heracles stream. The halo sample is represented with grey dots, the Heracles first dip population with semi-transparent blue circles and the second dip in red. Third column: the same plots are repeated for the Arnus I stream. The halo sample is represented with grey dots, the Arnus I population with semi-transparent blue circles.}
   \label{fig:ElementsGES}
\end{figure*}

Gaia-Enceladus is particularly well grouped within a broad dip of the reachability plot, and we observe that it appears to consist of two sub-groups (667-992 and 993-1438 in the ordered instances shown in Fig.~\ref{fig:stramreach}). This division is consistent and occurs every time we apply the learning algorithm, rather than being a one-time occurrence.
To investigate the differences between the two dips of GES, we plot them in both the Lindblad diagram (first column, first row) and the [$\alpha$/M] versus [M/H] plane (first column, second row) in Fig.~\ref{fig:ElementsGES}. We observe that the first dip corresponds to stars with lower energy and contains a significant number of stars excluded from the selection by \citet{Horta2023}. The energy distributions of the two dips peak at different values. In the first column, second row of Fig.~\ref{fig:ElementsGES}, we see that the mean abundances of $\alpha$-elements are slightly higher for stars in the first dip of the reachability plot.
We then examined the abundances of other elements to determine whether they reveal two distinct chemical structures. Specifically, we analysed the abundances of Mg, Al, Ce, Mn, Ni, and Na for the two GES substructures. We find that $\alpha$-elements and Al are fairly consistently separated, with the first dip showing higher abundances than the second. In contrast, the abundances of other elements appear more homogeneous across the two dips. This likely reflects the fact that {\sc Creek} used $\alpha$-elements and [M/H] to distinguish the {\sc Optics} groups.
The difference between the two sub-groups was tested using the two-sample Kolmogorov-Smirnov test \citep{Kolmogorovan1933sulla}, a non-parametric test that determines whether two samples originate from the same distribution. We assume the null hypothesis that the two samples—abundances for the two sub-groups—come from the same distribution. The significance level is set to 0.05, corresponding to a 5\% probability of rejecting the null hypothesis when it is actually true. We find some evidence that the two sub-groups may be originating from separate distributions for [M/H], [Mg/Fe], [Ni/Fe], [Ce/Fe], [O/Fe], [Si/Fe], and [Al/Fe]. However, they are from the same distribution for [Mn/Fe], [Na/Fe], and [Cr/Fe]. Therefore, what appears as a single structure in the selection by \citet{Horta2023} is likely more extended in energy, with slightly different abundances (or a gradient in abundances), as shown in Fig~\ref{fig:ElementsGES}. However, when considering the typical abundance uncertainties, as shown in Fig.~\ref{fig:ElementsGES}, the differences between the two subgroups might be considered minimal, with abundances agreeing within 1$\sigma$. 
Our test nevertheless demonstrates that the literature selection adopted by GES can be extended to include a larger number of stars with consistent chemical and dynamical properties.

\paragraph{Heracles (Kraken)}
Another interesting stream is Heracles, shown in green in Fig.~\ref{fig:stramreach}, situated between the ordered instances 2315-2399. This stream is well separated into two dips, although the energy and chemical properties appear more intertwined, as seen in the second column of Fig.~\ref{fig:ElementsGES}. Additionally, it seems that {\sc Creek} identifies a smaller percentage of stars for Heracles compared to GES, with some stars grouped in the dip around 2800. As discussed for the two populations in GES, the inherent uncertainties in the abundances prevent us from confirming their true existence.

\paragraph{Thamnos, Sequoia, Arjuna, and Helmi streams}

These streams are composed of fewer members in the selection by \citet{Horta2023} and do not appear to be clustered in either chemical or dynamic spaces. For instance, some stars of Thamnos fall within one of the dips around 3200, while others are scattered in more distant regions. Similarly, some stars from Sequoia, as identified in the Myeong and Naidu selections, fall into a dip around 1800, with the remainder dispersed across different dips. However, the existence of these streams remains uncertain in other recent studies as well \citep{dodd2023gaia, ceccarelli2024comparative}.

\paragraph{A new candidate stream: Arnus~I}

By observing the most significant dips in the reachability plot, we identify another interesting group. Although this group does not intersect with any known globular clusters or streams, it contains some members of Sequoia. As highlighted by the vertical band in pink in Fig.~\ref{fig:stramreach}, this group is located between instances 1874-1931. The dip exhibits well-clustered abundances and energies, particularly in the central region of the Lindblad diagram (third column, first row of Fig.\ref{fig:ElementsGES}). Its distribution is similar to that of the empirical selection of Sequoia members \citep[see e.g.][]{massari2019origin, Horta2023}, but it was obtained in a fully automatic and consistent manner, considering both kinematic and chemical properties. It shows some similarities also with Erebus, as identified by \citet{giribaldi2023chronology}. Notably, it exhibits chemical properties typical of in situ structures, a retrograde motion, and a metallicity spread comparable to that of the MW. However, the two structures do not have known stars in common. This could result from selection biases inherent to the different surveys employed (GALAH for \citet{giribaldi2023chronology} and APOGEE for the present work), but it does not constitute evidence that they represent the same structure. Abundances for some elements are shown in the third column, third, fourth and fifth row in Fig.~\ref{fig:ElementsGES}. We designate this group of stars as Arnus~I.

\section{Summary and conclusions}
\label{Section5}

Among the several populations of the Galaxy, the halo is the one that presents the most favourable conditions for attempting to recover destroyed structures.
Recovering disrupted mergers with clusters and dwarf satellites in the MW halo is essential for unravelling its complex merging history, for understanding the origin of the current structure of our Galaxy, and shedding light on fundamental questions about the formation and evolution of galaxies.
However, despite the ongoing efforts to identify large-scale streams conducted using astrometry and kinematics \citep[see e.g.][]{shih2024via}, the characterisation of nearby streams has proven even more complicated. Often, as in the case of GES, nearby streams or debris compose a large part of the local stellar halo, thus making their spatial identification difficult \citep{helmi2018merger}. Therefore, they can only be identified by the combined use of some of their invariant properties, such as orbital energy and angular momentum or actions and chemical composition.

In this work, we have presented a thoughtful attempt to make use of both the orbital and chemical properties of a selected sample of halo stars observed by APOGEE.
We outlined our algorithm, {\sc Creek}, which allowed us to take into account orbital parameters and chemical composition to recover star clusters and to identify or re-identify stellar streams, mergers, and debris. The algorithm consists of a combination of NNs, which produce a latent space in which {\sc Optics} is applied. 
We applied {\sc Creek} to a selection of APOGEE halo stars. 
In particular, the recovery of groups was quite effective once comparison to real globular clusters was made, with approximately 80\% of clusters being recovered with a 30\% threshold of completeness and homogeneity.
We were able to verify some of the obtained groups by comparing our results with known literature streams and merger remnants, for example GES and Heracles, and to identify a possible new one, a group of retrograde stars with chemical abundances typical of in situ populations named Arnus~I. 
In addition, in the identification of members of GES, {\sc Creek} was able to separate two substructures within it that are different both kinematically and chemically. The difference is small and falls within the error margins, yet it enables the automatic extension of the GES population.

With our method, we have established a repeatable, objective, and automatic technique for identifying and characterising streams and merger remnants. The limitations of the current databases, related to the low number of halo stars and the large uncertainties on the chemical abundances, currently restrict the results of our analysis. Moreover, in future applications, uncertainties on the orbital parameters could be addressed by sampling multiple orbital realisations for each star, thus allowing the model to be trained on uncertainty-aware data.
Methods similar to {\sc CREEK}, \citep[see e.g.][]{chen2018chemodynamical} applied to both field stars and globular clusters, yielded a modest recovery performance, identifying only four of nine known clusters, and just three exceeded a 64\% recovery fraction (see their Table 1), even with kinematic data. Our results show that graph autoencoders improve chemical tagging performance and allow for the detection of chemically coherent groups within a realistic field star background—which is a key advancement in validating chemical tagging for Galactic archaeology.
In the near future, there will be opportunities to apply {\sc Creek} to larger datasets with more precise abundances thanks to large spectroscopic surveys such as those carried out with WEAVE \citep{Jin2023MNRAS.tmp..715J} and 4MOST \citep{deJong2019Msngr.175....3D} and to forthcoming telescopes such as WST \citep{WST2024arXiv240305398M} that will provide precise abundances for a large set of halo stars covering many nucleosynthetic channels. For the current dataset of 3548 stars, running {\sc Creek} on a standard CPU-only laptop took approximately ten minutes for the GNN training and a few seconds for the OPTICS clustering. Future surveys such as 4MOST and WST are expected to provide significantly larger datasets. However, we are confident our approach is scalable because NN components are inherently capable of handling large data volumes, especially if they can exploit a GPU, and OPTICS, which we used here only for its visualisation properties, can be substituted by faster and equally effective density-based clustering algorithms such as Hierarchical Density-Based Spatial Clustering of Applications with Noise (HDBSCAN).

\begin{acknowledgements}
L.B., L.M.,   J.S.U., and R.E.G. thank INAF for the support (Large Grants EPOCH and WST), the Mini-Grants Checs (1.05.23.04.02), and the financial support under the National Recovery and Resilience Plan (NRRP), Mission 4, Component 2, Investment 1.1, Call for tender No. 104 published on 2.2.2022 by the Italian Ministry of University and Research (MUR), funded by the European Union – NextGenerationEU – Project ‘Cosmic POT’ Grant Assignment Decree No. 2022X4TM3H by the Italian Ministry of the University and Research (MUR). J.S.U. thanks INAF for its support through the Mini-Grant (1.05.24.07.02).
We thank Daniel Horta for providing his stream dataset from APOGEE and for his prompt and kind response, Carlos Viscasillas Vazquéz for his help in improving the quality of the figures, Rodolfo Smiljanic for his help and suggestions, and 
Sara Lucatello for the insightful discussion.
\end{acknowledgements}

\bibliographystyle{aa} 
\bibliography{aa55272-25corr.bib}

\end{document}